%% file: neurips_2024.tex
\documentclass{article}

\PassOptionsToPackage{numbers, compress}{natbib}

\usepackage[final]{neurips_2024_ml4ps}

\usepackage[utf8]{inputenc} 
\usepackage[T1]{fontenc}    
\usepackage{hyperref}       
\usepackage{url}            
\usepackage{booktabs}       
\usepackage{amsfonts}       
\usepackage{nicefrac}       
\usepackage{microtype}      
\usepackage{xcolor}         
\usepackage{graphicx}
\usepackage{amsmath}
\usepackage{physics}

\input{our_commands}

\title{Speak so a physicist can understand you!\\ TetrisCNN for detecting phase transitions \\and order parameters}

\author{%
  Kacper Cybiński\\
  Faculty of Physics, University of Warsaw, Pasteura 5, 02-093 Warsaw, Poland \\
  \And
  James Enouen\\
  Department of Computer Science, University of Southern California, \\Los Angeles, CA 90089, USA \\
  \And
  Antoine Georges \\
  Center for Computational Quantum Physics, Flatiron Institute, \\162 Fifth Avenue, New York, NY 10010, USA \\
  Collège de France, 11 place Marcelin Berthelot, 75005 Paris, France \\
  CPHT, CNRS, École Polytechnique, IP Paris, F-91128 Palaiseau, France \\
  DQMP, Université de Genève, 24 quai Ernest Ansermet, CH-1211 Genève, Switzerland \\
  \And
  Anna Dawid\thanks{Corresponding author.} \\
  Center for Computational Quantum Physics, Flatiron Institute, \\162 Fifth Avenue, New York, NY 10010, USA\\
  \texttt{adawid@flatironinstitute.org}
}

\begin{document}

\maketitle

\begin{abstract}
Recently, neural networks (NNs) have become a powerful tool for detecting quantum phases of matter. Unfortunately, NNs are black boxes and only identify phases without elucidating their properties. Novel physics benefits most from insights about phases, traditionally extracted in spin systems using spin correlators. 
Here, we combine two approaches and design TetrisCNN, a convolutional NN with parallel branches using different kernels that detects the phases of spin systems and expresses their essential descriptors, called order parameters, in a symbolic form based on spin correlators. We demonstrate this on the example of snapshots of the one-dimensional transverse-field Ising model taken in various bases. We show also that TetrisCNN can detect more complex order parameters using the example of two-dimensional Ising gauge theory. This work can lead to the integration of NNs with quantum simulators to study new exotic phases of matter.

\end{abstract}

\section{Introduction}

Machine learning promises a revolution in quantum sciences \cite{carrasquilla_machine_2020, Krenn23MLforQT, dawid2023modern, medvidovic2024NQS}, similar to the current revolution in industry~\cite{dawid2023introduction}. Recently, neural networks (NNs) have become a powerful tool for detecting phases of matter \cite{Carrasquilla17NatPhys, van_nieuwenburg_learning_2017, wetzel_unsupervised_2017, greplova_unsupervised_2020, bohrdt2021analyzing, patel2022unsupervised, arnold2024generative, kim2024attention}, which is especially promising in the context of experimental data \cite{Rem19, Khatami20, Kaming2021, link2023machine, miles2023machine, sadoune2024learnSPTexp}. An ultimate goal of this forefront is to make such automated approaches interpretable \cite{Dawid20NJP, wetzel_discovering_2020, Dawid2021Hessian, arnold_interpretable_2021, arnold2022replacing, arnold2023fisher, wetzel2024closedform, zhan2024learning, cybinski2024ssh}, which should lead to learning phases of matter and understanding their properties, in particular the detected order parameters~\cite{miles2023machine, sadoune2024learnSPTexp, wetzel2017orderparams, Greitemann19probing, Liu19interpretable, Greitemann19identification, sadoune2023unsupinterpet, sadoune2024humanmachine, cole_quantitative_2021, Miles2021CCNN, schlomer2023fluctuationinterpret, cao2024unveiling, suresh2024ctransformer}.

Here, we report on designing a convolutional neural network (CNN) with parallel convolutional branches that use kernels of different sizes and shapes. We dub this network `TetrisCNN` and show here that it detects the phases of spin systems and expresses their order parameters in a symbolic form based on spin correlators. TetrisCNN takes inspiration from \reflabs~\cite{wetzel2017orderparams, Miles2021CCNN} and improves on them in terms of simplicity, computational cost, and inclusion of multiple measurement bases. Moreover, we use symbolic regression (SR)~\cite{cranmer2023pysr} to provide symbolic formulas for the detected order parameters and the network itself, making our method particularly attractive to physicists.

\begin{figure}[t]
    \centering
    \includegraphics[width=\columnwidth]{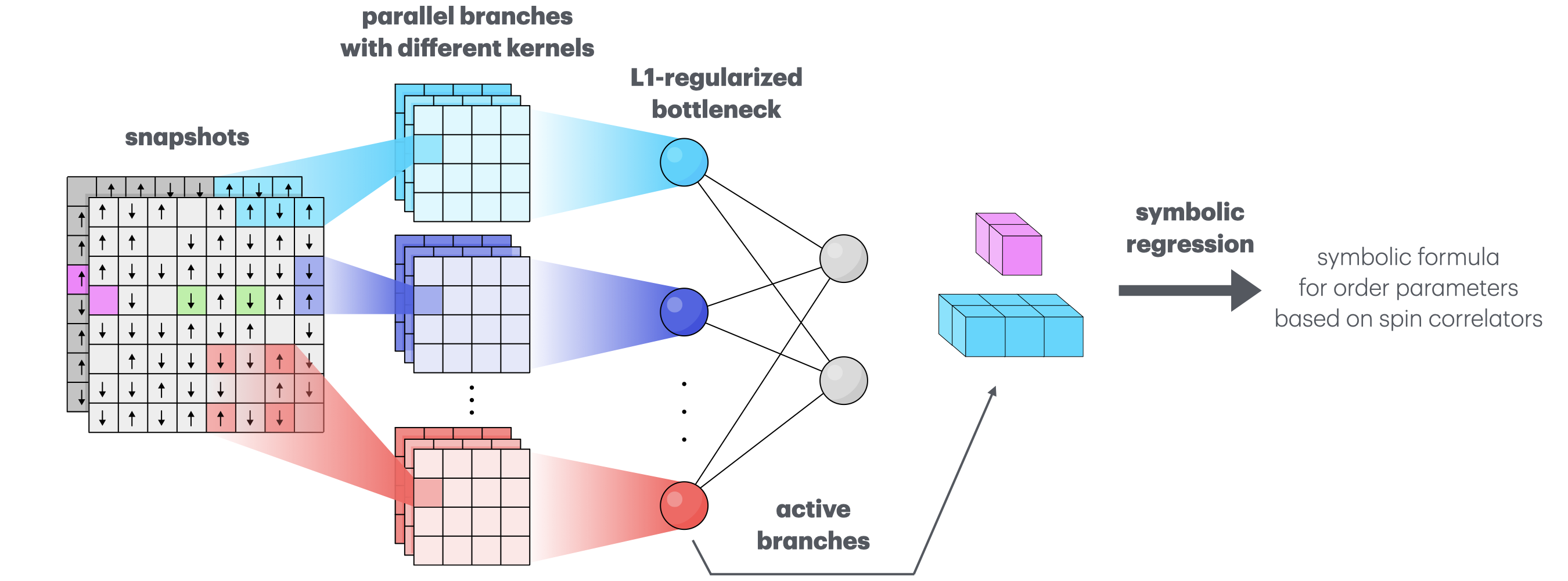}
    \caption{\textbf{TetrisCNN is composed of parallel branches that use kernels of different shapes that map to spin correlators.} Its regularized bottleneck suppresses activations of branches containing non-unique or irrelevant information. The remaining active branches are analyzed with symbolic regression (SR), which provide a symbolic formula for the network based on spin correlators.}
    \label{fig:intro}
\end{figure}

\section{Methods}\label{sec:methods}

We test TetrisCNN ability to detect phase transitions and relevant order parameters on the example of three datasets derived from two paradigmatic spin models: 1D quantum transverse-field Ising model (TFIM) and 2D Ising gauge theory (IGT). Each spin at site $i$ can take only two values $S_i = \pm 1$. We use $\overline{S_i} = \frac{1}{N} \sum_i S_i$ to symbolize averaging over all $N$ system sites.

\paragraph{1D transverse-field Ising model (TFIM).} 1D TFIM describes a spin chain of length $N$ that exhibits quantum phases. The Hamiltonian of the system is given by:
\begin{equation}\label{eq:TFIM_ham}
    \hat{H}_{\rm TFIM} = -J ( \sum_{i} \hat{S}^z_i \hat{S}^z_{i + 1} + g \sum_i \hat{S}^x_i )
\end{equation}
where $J$ is the spin-spin interaction strength, $g$ is the transverse field strength, and $\hat{S}^z$ and $\hat{S}^x$ are Pauli matrices representing the spin operators. When $J$ dominates $g$, the system is in the ordered ferromagnetic phase with all spins aligned with each other in the $z$ direction. The critical point of the system occurs when $g$ balances $J$, and the system transitions to the disordered (paramagnetic) phase. 
We obtain ground states of 1D TFIM for various values of $g$ ($J = 1$, $N=150$) using the Density Matrix Renormalization Group (DMRG) \cite{white1992dmrg} and ITensors.jl \cite{stoudenmire2016tensornetworks}. We take measurements (snapshots) of ground states in $z$ and $y$ bases and use them as input data for TetrisCNN to imitate experimental measurements on a quantum system.

\paragraph{2D Ising gauge theory (IGT) model} IGT is a classical spin model that exhibits a topological phase of matter, defined on a square lattice with periodic boundary conditions. This time the interaction between spins is defined within plaquettes $p$ on the lattice and is of four-body nature:
\begin{equation}\label{eq:IGT_ham}
    H_{\rm IGT} = - J \sum_{p} \prod_{i \in p} S_i\,.
\end{equation}
The (highly degenerate) ground state of this Hamiltonian meets the local constraint that the product of spins along each plaquette is $\prod_{i \in p} S_{i}=1$. The system exhibits a transition from the low-temperature topological phase to the high-temperature phase with violated constraints (see \seclab~3.1.2 in Ref.~\cite{dawid2023modern}). We obtain spin configurations of IGT for different temperatures using the Monte Carlo method~\cite{greplova_unsupervised_2020}.

\paragraph{Architecture of TetrisCNN}
The main idea behind TetrisCNN is to allow the network to utilize a multitude of differently shaped convolutional kernels, similar to Tetris pieces, within multiple parallel branches, as presented in \fig~\ref{fig:intro}. The results of their parallel computation, after global average pooling, enter the TetrisCNN bottleneck as $a_k$, which we regularize with $ L_{\rm bottle} =  \sum_k \lambda_k | a_k |$, where $k$ is a branch index. As such, we deactivate branches without important and unique information. We additionally choose $\lambda_k$ so that they promote the use of simpler kernels over more complicated ones. After the bottleneck, fully-connected layers solve the posed task, given the learned sparse data representation. We describe the architecture in more detail in \app~\ref{app:detailed_archi}.

\paragraph{TetrisCNN interpretation}
If we apply TetrisCNN now to spin configurations (where each input element can be only $S_i = \pm 1$), due to TetrisCNN construction, each branch activation $a_k$ can be mapped exactly to a linear function of a specific spin correlator (see \app~\ref{app:ss-archi} and \reflab~\cite{wetzel2017orderparams} for more details). For example, the branch using kernel (2,1) can only learn a nearest-neighbor spin correlator $\overline{S_i S_{i+1}}$ (it can also learn a one-body $\overline{S_i}$ but we can ignore it if the branch using (1,1) got deactivated). Therefore, we can easily understand what each branch computes, and linear regression is enough to find a formula for $a_k$. Thanks to the bottleneck sparsity, we can also obtain a symbolic formula for the whole network using SR \cite{cranmer2023pysr} (see \app~\ref{sec:app:interpretation_theory}), which otherwise fails due to the large input size.

\paragraph{Task} In the following, we combine TetrisCNN with the prediction-based method \cite{schafer_vector_2019, greplova_unsupervised_2020}, which allows for an unsupervised detection of phase transitions. First, the network is trained in a regression task to predict a tuning parameter of the system. For 1D TFIM, it is the transverse field value, $g$; for 2D IGT, it is an inverse temperature, $\beta$. Then, the phase transition is identified by locating the maximum of the derivative of the network output as a function of the label. Intuitively, the prediction-based method relies on the difficulty of a successful regression in the transition vicinity. We provide the implementation details in \app~\ref{app:numerical_details} and in the accompanying GitHub repository~\cite{OurRepo}.

\section{Results}\label{sec:results}

\begin{figure}
    \centering
    \includegraphics[width=\columnwidth]{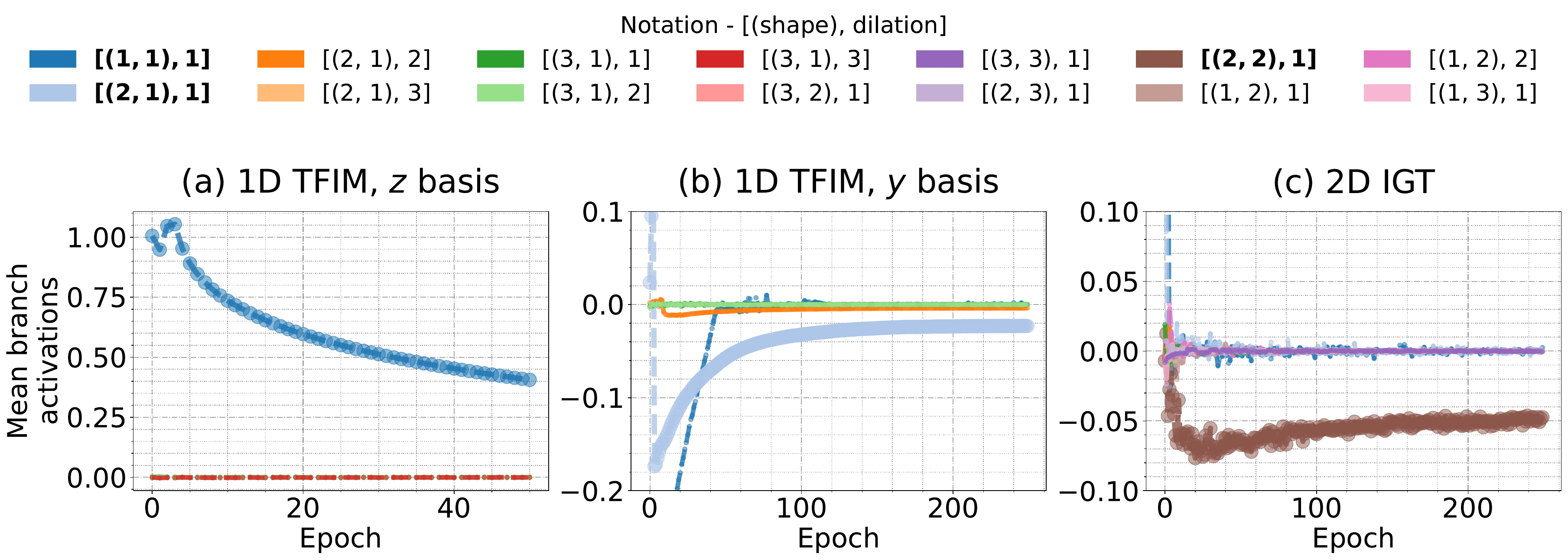}
    \caption{\textbf{Activation of TetrisCNN branches across training on different datasets}. During training, branches without uniquely important information die out. In all three datasets, only one branch remains active. It relies on (a) the (1,1) kernel in the 1D TFIM measured in $z$ basis, (b) the (2,1) kernel in the 1D TFIM measured in $y$ basis, and (c) (2,2) kernel in the 2D IGT.}
    \label{fig:kernels_vs_epochs}
\end{figure}

\paragraph{The branch activation depends on relevant correlators present in the data}

The interpretability of TetrisCNN relies on the sparsity of its bottleneck, i.e., a small number of its non-zero elements. Ideally, the bottleneck should contain only crucial information on the dataset. In \fig~\ref{fig:kernels_vs_epochs}, we show how, during training, TetrisCNN branches that pick up irrelevant correlations in the data deactivate, i.e., their respective activations in the bottleneck go to zero. The branch (or branches) that remains active uses the kernel that computes the relevant correlators present in the data. 

For 1D TFIM, the order parameter that determines the transition between the ferro- and paramagnetic phases is the system magnetization in the $z$ direction, $\langle \overline{\hat{S}_i^z} \rangle$. It is a one-body quantity, and as expected, in \fig~\ref{fig:kernels_vs_epochs}(a) we see that the remaining active branch of TetrisCNN uses the kernel of shape (1,1). However, the magnetization in the $y$ direction does not have information on the phase transition. In \fig~\ref{fig:kernels_vs_epochs}(b) we see that when TetrisCNN analyzes TFIM measurements taken in $y$ basis, it uses the branch with the kernel of shape (2,1). Indeed, we can check that the value of the nearest-neighbor spin correlator $\langle \overline{\hat{S}_i^y \hat{S}_{i+1}^y} \rangle$ changes fast when the system undergoes a phase transition. Finally, TetrisCNN correctly focuses on the relevant correlator of the IGT, i.e., a four-body spin correlator related to the plaquette. We study how the detected correlators depend on  $\lambda_k$ in $L_{\rm bottle}$ in \app~\ref{app:lambda_sparsity}.

\begin{figure}
    \centering
    \includegraphics[width=\columnwidth, trim={0 5cm 0 0}, clip]{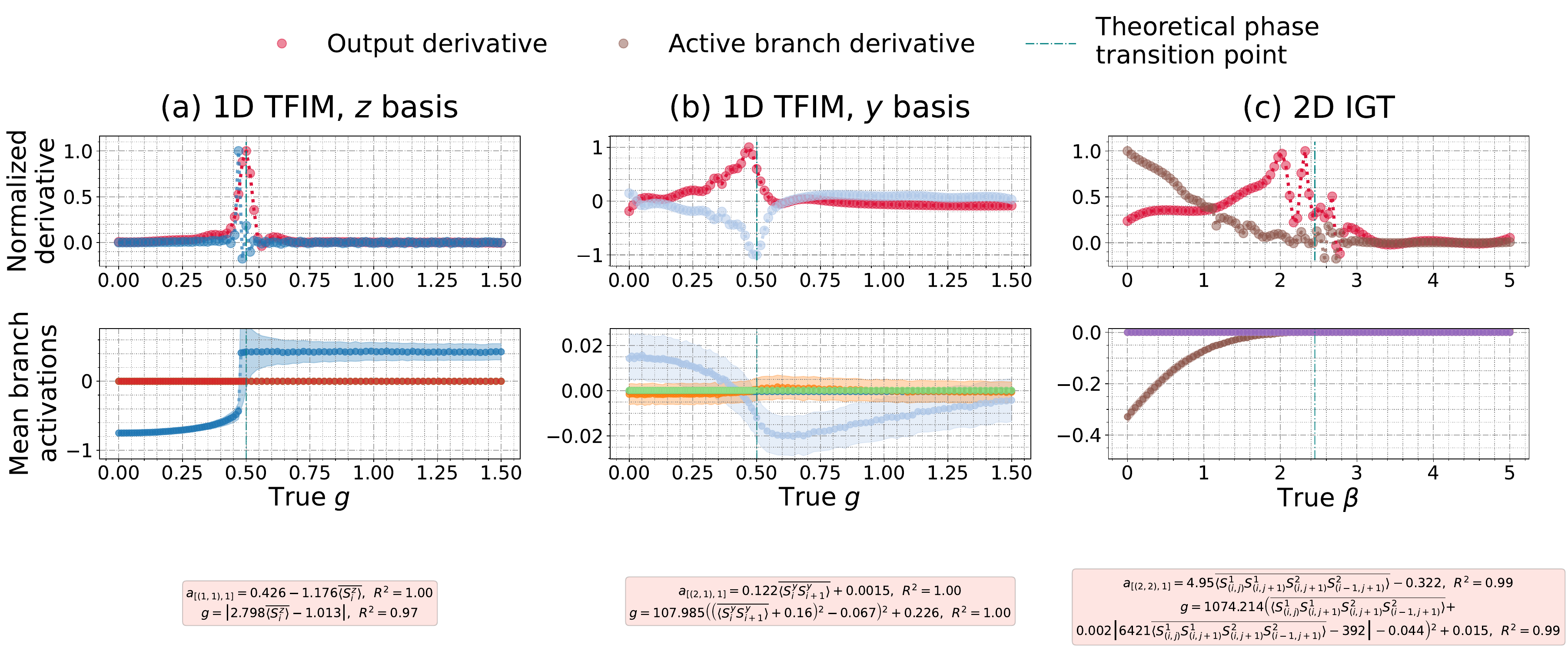}
    \caption{\textbf{Identification of phase transition location and the relevant spin correlators with TetrisCNN and prediction-based method.} (a)-(b) Results for 1D TFIM measures in $z$ and $y$ basis, respectively. (c) Results for 2D IGT.}
    \label{fig:prediction_and_order_param}
\end{figure}

\paragraph{TetrisCNN identifies the phase transition location in an unsupervised way}


In the first row of \fig~\ref{fig:prediction_and_order_param}, we recover unsupervised detection of phases from \reflabs~\cite{schafer_vector_2019, greplova_unsupervised_2020}. The predicted transition locations are in good agreement with the theoretical predictions for both Ising models. 
In the second row of \fig~\ref{fig:prediction_and_order_param}, we plot the branch activations as functions of the respective label, i.e., $g$ for 1D TFIM and $\beta$ for 2D IGT. We see that the values of the active branches exhibit a fast change in the vicinity of the phase transition, reminding of the expected behavior of the order parameter. As a result, we can also recover the transition location by studying the derivative of the branch activation for data for different labels. The maxima of the derivatives of the branch activations and of the network output are slightly shifted, indicating different locations for the phase transition. The reason for this will become clear in the next section. 

\paragraph{TetrisCNN identifies the order parameter and provides its symbolic formula}

Most excitingly, we can analyze the trained TetrisCNN with SR. For brevity, we focus here on results for TetrisCNN trained on 1D TFIM in $y$ basis and place the rest in \app~\ref{app:sr_results}.
The first mapping we find is between the dominant branch activation and the spin correlation it detects. As seen in \fig~\ref{fig:kernels_vs_epochs}(b), TetrisCNN has an active branch that uses [(2,1),1] kernel and its activation is
\begin{equation}
    a_{[(2,1),1]} = 0.122 \langle \overline{ S_i^y S_{i+1}^y} \rangle + 0.0015 \,\, {\scriptstyle(R^2 = 1.00)}\,,
\end{equation}
where $\langle \cdot \rangle$ means averaging over snapshots or spin configurations obtained for the same tuning parameter. $R^2$ coefficient in parenthesis shows that the fit is excellent. We see that in 1D TFIM, the active branches of TetrisCNN learn the magnetization in the $z$ direction from the data measured in the $z$ basis and a nearest-neighbor spin correlator in $y$ direction from the data measured in the $y$ basis. In 2D IGT, TetrisCNN branch is a linear function of the plaquette $\langle \overline{ S_{(i,j)}^1 S_{(i,j+1)}^1 S_{(i,j+1)}^2 S_{(i-1,j+1)}^2} \rangle$. 

The next step is to find the mapping between the non-zero branch activations and the network output. In the same example, the predicted $g$ is
\begin{equation}
    g = 107.985 \left( \left( \langle \overline{ S_i^y S_{i+1}^y} \rangle + 0.16 \right)^2 - 0.067 \right)^2 + 0.226  \,\, {\scriptstyle(R^2 = 1.00)}\,.
\end{equation}
As we see, the full network is a cumulant of the detected spin correlator. This explains the shift in the phase transition location predicted from the derivative of the branch activation compared to the derivative of the network output. We leave the question of which is a better indicator of the phase transition for future work.

\section{Limitations and Outlook}

An obvious limitation of TetrisCNN is the combinatorial complexity of kernels that increases fast for highly non-local order parameters. Currently, it can handle correlators up to (5,5). However, the main challenge is at the level of finding a symbolic formula for the whole network, while interpreting the bottleneck scales much better thanks to the direct linear mapping between the kernel shapes and spin correlators. In a way, TetrisCNN automatically searches for relevant correlators \cite{verdel24datadriven}, but also learns an arbitrary useful function of correlators. Moreover, the learned correlators should be task-dependent, but we leave this aspect for future work.

The next steps are to expand TetrisCNN to spin models on lattices of different geometries. An exciting avenue is to modify the approach to detect and give formulas for long-range, nematic, and topological orders in data, following the developments of \reflabs~\cite{sadoune2024learnSPTexp, suresh2024ctransformer}.
Although TetrisCNN was developed with spin models in mind, it may tackle other correlation-based tasks, e.g., in molecular gases.

\ack{We thank Eliška Greplová for useful discussions. K.C. acknowledges the financial support from the Polish Ministry of Science and Higher Education within the ``Excellence initiative – research university'' program. The Flatiron Institute is a division of the Simons Foundation.}

\bibliographystyle{apsrev4-2.bst}
{\small \bibliography{bibliography}}


\appendix

\section{Detailed explanation of the TetrisCNN architecture}\label{app:detailed_archi}

\subsection{TetrisCNN architecture}\label{app:ss-archi}

\begin{figure}[t]
    \centering
    \includegraphics[width=\linewidth]{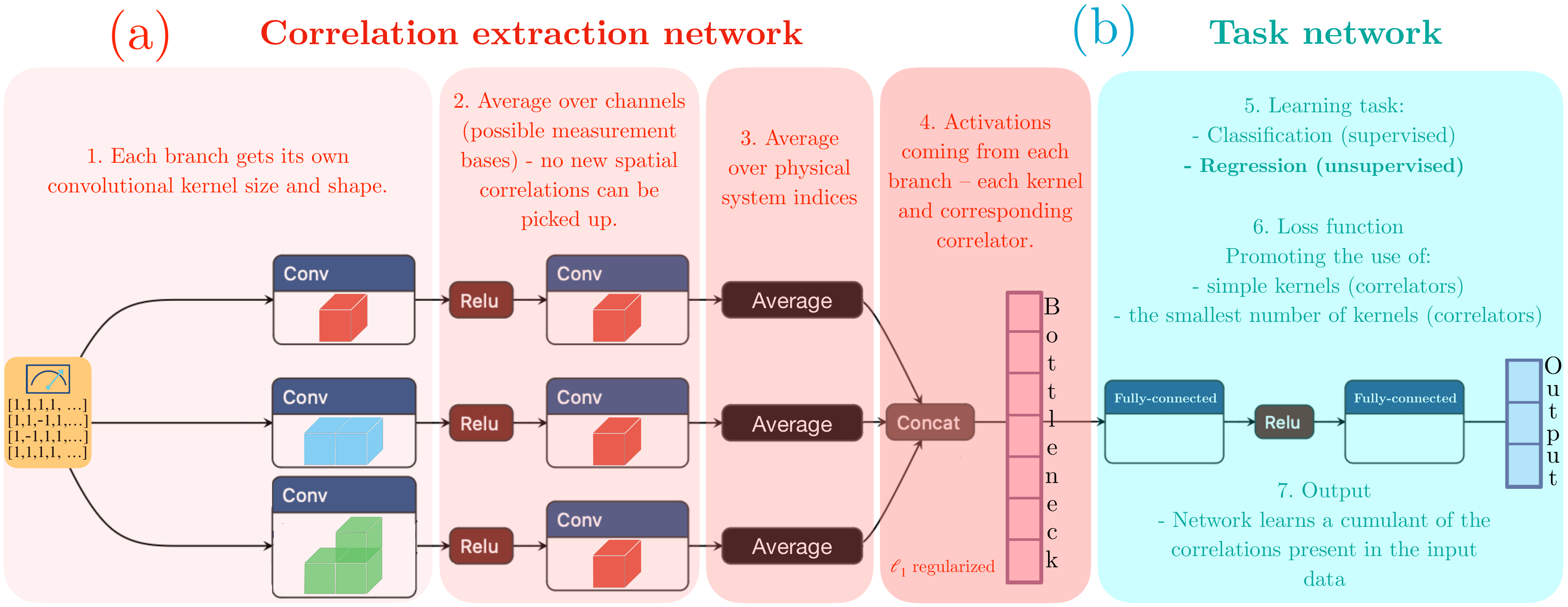}
    \caption{\textbf{A detailed presentation of TetrisCNN architecture and how it processes spin configurations at each step.} (a) The first part of TetrisCNN, the correlation extraction network, which maps the spin configurations onto functions of many-body spin correlators present in the data. (b) The second TetrisCNN subnetwork is the task network, in our case a fully-connected network. If applied to phase detection problem, the task can bebe either unsupervised (as here, where we use the prediction-based method) or supervised.}
    \label{fig:tetris_detailed}
\end{figure}

The TetrisCNN architecture can be divided into two general parts: correlation extraction and task networks. They are presented in \fig~\ref{fig:tetris_detailed}(a)-(b), respectively, along with a detailed description of each data processing step. We discuss these steps in the following sections.

\textbf{Extracting correlations.} Let us first focus on the correlation extraction network. The input to the network are either classical spin configurations or 'snapshots' - projective measurements on a simulated or experimentally realized quantum spin lattice. The architecture supports either one- or two-dimensional systems with snapshots in any of $x, y, z$ measurement bases or their combination. Each combination of input bases results in a separate training and analysis routines. The snapshots are introduced into the network and then processed in parallel by several 'branches', reminiscent of bilinear CNNs from \reflab~\cite{lin2015bilinear}. Each branch consists of a convolutional layer and then averages over the number of filters and physical indices, thus reducing each snapshot to a single number - a branch activation $a_k$. 
Following the logic presented by Ref.~\cite{wetzel_unsupervised_2017}, we interpret this activation as a function of all correlators present in the data of the same shape and size or smaller as the convolutional kernel used in the branch. Therefore, each activation $a_k$ can be written as a function of the input snapshot - set of spin variables $S_i$:
\begin{equation}
    a_k = f(\qty{S_i^z}) 
\end{equation}
We map the spin variables from $\qty{\uparrow, \downarrow}$ to $\qty{-1, 1}$. Then, further following the logic from Ref.~\cite{wetzel2017orderparams}, we leverage the fact that:
\begin{equation}
    (S_i^d)^p = \begin{cases}
    1 \quad &\mathrm{if} \,\, p \,\, \text{is even}\\
    S_i^d \quad &\mathrm{if} \,\, p \,\, \text{is odd}
    \end{cases},
\end{equation}
and apply it to a Taylor expansion of the function:
\begin{equation}
    f\left(S_i^z\right)=f_0 +f_1S_i^z+f_1\underbrace{\left(S_i^z\right)^2}_{1}+f_2\underbrace{\left(S_i^z\right)^3}_{S_i^z}+\ldots
\end{equation}
This leaves us with an activation $a_k$ being a function of all correlations of smaller or equal size than the kernel size, but only of the first order. An example of this for kernels (1,1) and (2,1) would be:
\begin{align}\label{eq:1_1_z}
    f\left(s_i^z\right) &= F\left(\frac{1}{N}\sum_i S^z_i\right)\\  \label{eq:2_1_y}
    f\left(s_i^y, s_{i+1}^y\right) &= F\left(\frac{1}{N}\sum_i S^y_i\right) \, + F\left(\frac{1}{N-1}\sum_i S^y_i S^y_{i+1}\right)
\end{align}
\textbf{Truncation of search space.} This trick still leaves us with the branch activation dependent on the correlations also smaller than the kernel size. To tackle this, we add a term to the loss function so that the network learns during the training to use only the simplest non-trivial kernels to achieve this task. This is achieved by combining the activations from all branches, forming a network bottleneck. This bottleneck is $\ell_1$ regularized, and the values of the activations are combined into $L_{\rm bottle}$, which is then added to the final loss function. As a result, once the network is trained, we can inspect the values of the activations in the analysis routine, and the ones with the highest amplitude are perceived as the most important to the network. Therefore, e.g., if we allow the network to be used in the optimization process, the kernels (1,1), (2,1), (3,1), and kernel (2,1) turns out to be the one with the highest amplitude, the function describing it presented in \eq~\eqref{eq:2_1_y} collapses only to the second term. The first term is either zero or not of the leading order because if it were of the leading order, then a kernel (1,1) would not be deactivated. Therefore, only by studying the branch activations, we can detect leading correlations in the data, solely from the interpretable design of TetrisCNN architecture.

\textbf{Learning task.} The second part of the network is the task network, which learns the mapping between the branch activations and the desired output. We implement two possible tasks: classification and regression. The variant with regression task is based on the prediction-based method~\cite{schafer_vector_2019, greplova_unsupervised_2020}, which allows for an unsupervised detection of the phase transition point, as opposed to the classification task, which requires the user to supply the labels.

\subsection{Interpretation with symbolic regression}\label{sec:app:interpretation_theory}

\begin{figure}
    \centering
    \includegraphics[width=\linewidth]{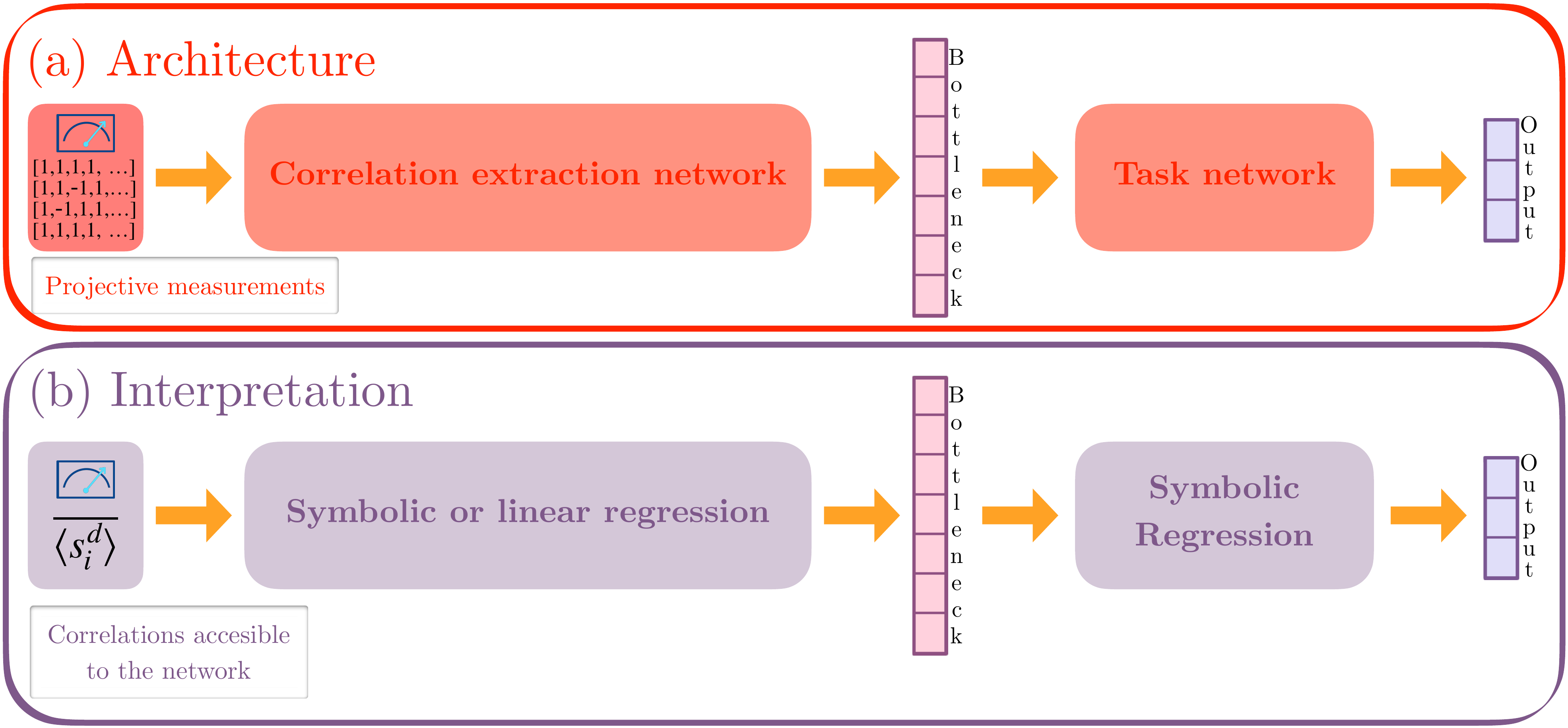}
    \caption{\textbf{A high-level overview of the TetrisCNN architecture and its interpretation.} The two parts form a pipeline for the unsupervised detection of phases of matter and their order parameters. (a) TetrisCNN architecture comprises two general parts - correlation extraction network and task network. The two are joined by a bottleneck, which is key to the subsequent analysis. (b) Obtaining symbolic formulas takes two steps. First, finding a mapping between correlations accessible to each branch and this branch activation with linear or symbolic regression. Second, finding a mapping between the network output and the branch activations in the bottleneck.}
    \label{fig:architecture_interpretation_high_level}
\end{figure}

Although the detection of relevant spin correlators can be done simply by studying the network bottleneck and leveraging its interpretable design, we can additionally use regression techniques to provide a symbolic formula for both bottleneck activations and the network output, all in terms of spin correlators. To this end, we first find a mapping between the spin correlators and the branch activations. Due to $\ell_1$ regularization of the bottleneck, increasing $\lambda_k$, and binary values of the spin data, the bottleneck elements $a_k$ can only be linear functions of the respective spin correlators, as explained above (i.e., kernel (1,1) can only learn $\overline{S_i}$, kernel (2,1) - only $\overline{S_i S_{i+1}}$). We can identify this mapping using linear regression (e.g., via least-squares minimization). However, to find a symbolic formula for the mapping between bottleneck activations $a_k$ and network output, we need to use symbolic regression (SR). These two steps are presented in \fig~\ref{fig:architecture_interpretation_high_level}(b).

\textbf{Symbolic regression.} Symbolic regression (SR) is a type of regression analysis that searches for mathematical expressions that best fit the given data rather than assuming a predefined functional form like in traditional regression. This approach is particularly useful in physics, where the goal is often to discover underlying equations or relationships that govern the observed data~\cite{cranmer2023pysr}.
The SR analysis was performed using the PySR package, a Python interface to the SymbolicRegression.jl library in Julia \cite{cranmer2023pysr}. This package employs evolutionary algorithms and genetic programming to iteratively search for and evolve mathematical expressions that best fit the activation data. Its procedure is the following:
\begin{enumerate}
    \item \textbf{Initialization:} The algorithm starts with a population of random mathematical expressions involving basic operators ($+\,,\,-,\,*,\,/$), constants, and the input variables (correlators).
    \item \textbf{Evaluation:} Each expression is evaluated based on its ability to predict the kernel activations. The evaluation metric is the default SymbolicRegression.jl loss function, which is the mean squared error (MSE) between the predicted and actual activations.
    \item \textbf{Selection and Evolution:} The top-performing expressions are selected, and new expressions are generated through genetic operations such as mutation (altering parts of an expression) and crossover (combining parts of two expressions).
    \item \textbf{Convergence:} Depending on the size of the input, the process continues for $6 000$ or $18 000$ epochs, which allows for a good convergence of each branch evolved by the genetic algorithm employed under the hood of PySR.
\end{enumerate}

\section{Details on the numerical implementation and used hyperparameters}\label{app:numerical_details}

\paragraph{Task} In this work, TetrisCNN is used in combination with the prediction-based method~\cite{schafer_vector_2019, greplova_unsupervised_2020}. We train TetrisCNN using PyTorch~\cite{PyTorch2} by minimizing the mean squared error between the network output and the tuning parameter. For 1D TFIM, the tuning parameter is the transverse field value, $g$; for 2D IGT, the tuning parameter is an inverse temperature, $\beta$. The training hyperparameters are in \tab~\ref{tab:hyperparams}.

\paragraph{Input data} The input data are spin configurations, either classical (2D IGT) or projective measurements taken on a quantum system in some bases (1D TFI). Single input data therefore is a vector (for 1D systems) or matrix of $\{-1,1\}$ (for 2D systems). If a system is classical or only one measurement basis is considered, input data has a single channel, as in case of the 1D TFI, where $z$ and $y$ bases were considered separately. If two measurement bases are considered, like in the 2D IGT case, the spin configurations (here from two sublattices) are randomly bunched into pairs consisting of two bases, and are fed into two input channels. For three measurement bases, there are three channels.

\paragraph{Declaring available branches} TetrisCNN requires giving a list of kernels that it needs to consider. The kernel format we use is $[({\rm dimension}_1, {\rm dimension}_2), {\rm dilation}]$, where ${\rm dimension}_2 = 1$ for 1D and dilation inserting holes of specified size between the kernel consecutive elements. Dilation $=1$ means no hole, dilation $=2$ means hole of one element size. Each kernel is then used by a separate parallel convolutional branch. In our numerical experiments, we used up to 20 parallel branches and did not notice any performance drop. We report the list of kernels that we made available for every setup described in the main text in \tab~\ref{tab:hyperparams}.

\paragraph{Bottleneck regularization} Interpretation of TetrisCNN via symbolic regression (SR) hinges upon the sparsity of its bottleneck. Therefore, we regularize the bottleneck elements $a_k$, i.e., branch activations averaged across channels and physical system size via the following loss term:
\begin{equation}
    L_{\rm bottle} = \lambda_k \sum_k | a_k | \,.
\end{equation}
This $\ell_1$ regularization encourages the bottleneck elements to go to zero, which translates into the use of the smallest number of branches (therefore kernels and spin correlators) possible. We additionally modulate the values of $\lambda_k$ so it is the smallest for branches using kernels of the smallest size, also favoring simpler correlations over more complex ones. We choose it so that they are equally distanced on a logaritmic scale between $\lambda_{\rm min}$ and $\lambda_{\rm max}$, $\lambda_ k =$ \texttt{np.logscale($\lambda_{\rm min}$, $\lambda_{\rm max}$)}.
We report values of $\lambda_{\rm min}$ and $\lambda_{\rm max}$ in \tab~\ref{tab:hyperparams}.

\begin{table}[]
\centering
\caption{Hyperparameters of TetrisCNN trained on three datasets, whose results are reported in the main text}
\label{tab:hyperparams}
\resizebox{\columnwidth}{!}{%
\begin{tabular}{|l|c|c|c|}
\hline
\textbf{}                              & \textbf{1D TFIM, z basis} & \textbf{1D TFIM, y basis}    & \textbf{2D IGT}              \\ \hline
\textbf{Learning rate}                 & 1e-2                      & 5e-4                         & 5e-2                         \\ \hline
\textbf{Weight decay}                  & 1e-2                      & 1e-1                         & 1e-5                         \\ \hline
\textbf{Optimizer}                     & Adagrad                   & AdamW                        & Adagrad                      \\ \hline
\textbf{Max epochs}                    & 100                       & 100                          & 1500                         \\ \hline
\textbf{Early stopping}                & Yes                       & No                           & No                           \\ \hline
\textbf{Normalized input}              & No                        & No                           & MaxMinScaler                 \\ \hline
\textbf{Available kernels} &
  \begin{tabular}[c]{@{}c@{}}{[}(1, 1), 1{]},\\ {[}(2, 1), 1{]},\\ {[}(2, 1), 2{]},\\ {[}(2, 1), 3{]},\\ {[}(3, 1), 1{]},\\ {[}(3, 1), 2{]},\\ {[}(3, 1), 3{]}\end{tabular} &
  \begin{tabular}[c]{@{}c@{}}{[}(1, 1), 1{]},\\ {[}(2, 1), 1{]},\\ {[}(2, 1), 2{]},\\ {[}(3, 1), 1{]},\\ {[}(3, 1), 2{]}\end{tabular} &
  \begin{tabular}[c]{@{}c@{}}{[}(1, 1), 1{]},\\ {[}(2, 1), 1{]},\\ {[}(1, 2), 1{]},\\ {[}(2, 2), 1{]},\\ {[}(2, 1), 2{]},\\ {[}(1, 2), 2{]},\\ {[}(3, 1), 1{]},\\ {[}(3, 2), 1{]},\\ {[}(1, 3), 1{]},\\ {[}(2, 3), 1{]},\\ {[}(3, 3), 1{]}\end{tabular} \\ \hline
\textbf{Bottleneck size (\# branches)} & 7                         & 5                            & 11                           \\ \hline
$\lambda_{\rm min}$, $\lambda_{\rm max}$                 & 1e-4, 1e0                 & 1e-4, 1e-1                   & 1e-5, 1e-1                   \\ \hline
\textbf{Task network}                  & {[}\# branches, 32, 1{]}  & {[}\# branches, 32, 16, 1{]} & {[}\# branches, 32, 16, 1{]} \\ \hline
\end{tabular}%
}
\end{table}

\paragraph{Architecture of the task network} While the correlation extraction network is the same across setups, we vary the fully-connected task network between the datasets. We report its architecture also in \tab~\ref{tab:hyperparams}, where the input to the task network is the bottleneck (therefore branch activations) and the next numbers indicate numbers of hidden neurons in the consecutive fully-connected layers.

\section{Interplay between the accuracy and sparsity of TetrisCNN via the bottleneck regularization}\label{app:lambda_sparsity}

\begin{figure}[t]
    \centering
    \includegraphics[width=\linewidth]{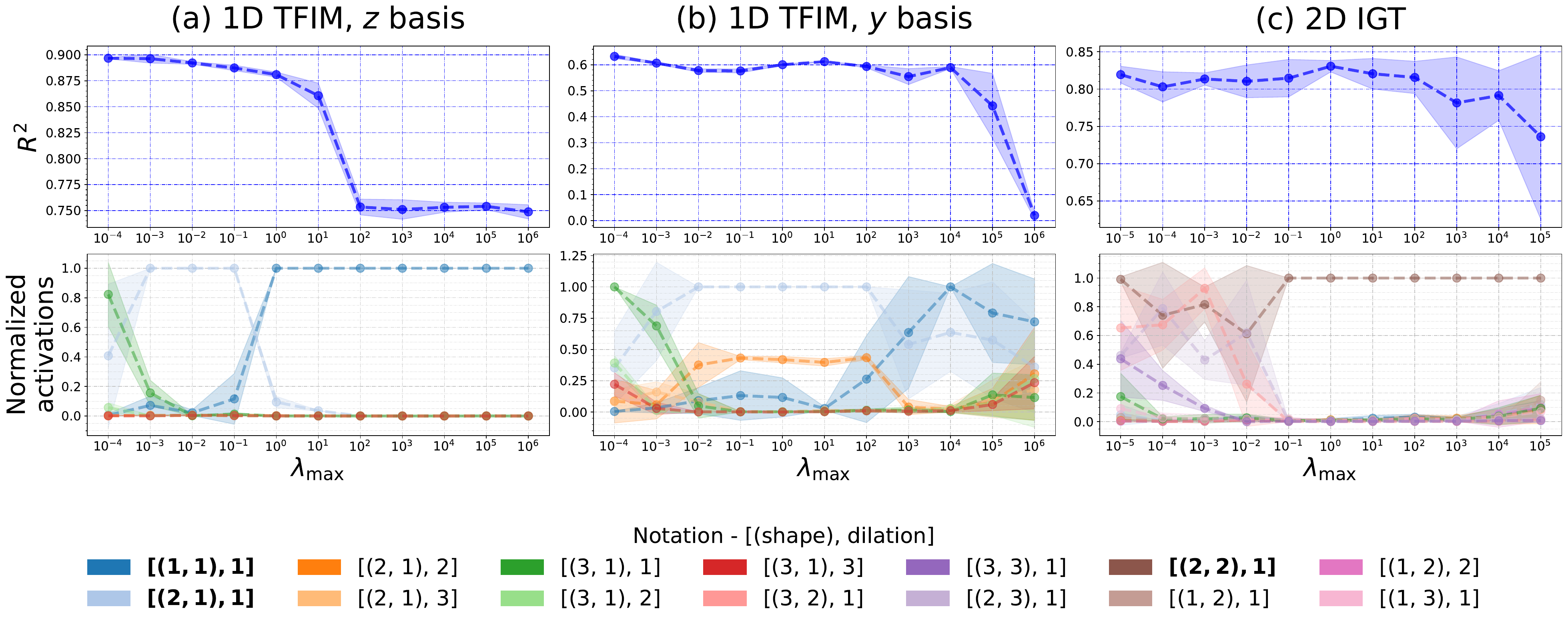}
    \caption{\textbf{Penalties of branch activations $\lambda_k$ and the use of kernels.} We plot here the average $R^2$ of the network output (compared to the true label) and the normalized branch activations (where 1 is the largest branch activation value) of 5 TetrisCNN instances trained with different $\lambda_{\rm max}$. The larger $\lambda_{\rm max}$, the larger differences between penalties $\lambda_k$ put on the branch activations that use kernels of increasing complexity.
    Increasing $\lambda_{\rm max}$ results in a sparser bottleneck of TetrisCNN that uses simpler and smaller kernels. At some critical $\lambda_{\rm max}$, we can also get a significant $R^2$ drop, so the sparsity may come at the cost of performance.}
    \label{fig:lambda_sparsity}
\end{figure}

Here we study how the detected spin correlators (and kernels used by the active branches of TetrisCNN) depend on penalties $\lambda_k$ put on the branch activations that use kernels of increasing complexity. We set $\lambda_k = $ \texttt{np.logscale($\lambda_{\rm min}$, $\lambda_{\rm max}$)}, where $\lambda_{\rm min}$ is fixed to $10^{-4}$ and $10^{-5}$ for 1D TFIM and 2D IGT, respectively. They are applied to the branch activations $a_k$ in the order of appearance on the list of the respective 'available kernels' defined by a user (see \tab~\ref{tab:hyperparams}). In \fig~\ref{fig:lambda_sparsity}, in the first row, we plot the average regression performance in terms of $R^2$ of 5 TetrisCNNs trained with increasing $\lambda_{\rm max}$ that is with increasing differences between applied $\lambda_k$ that put increasing preference towards smaller and simpler kernels (i.e., kernels from the beginning of the user-defined list). In the second row, we plot average branch activations of the same TetrisCNNs, which we normalize to the largest branch activation per network instance. We show here which kernels are preferred by differently regularized networks and also how stable is their choice across 5 initializations.

\paragraph{Uniform $\lambda_k$} When $\ell_1$ penalties put on the branch activations corrresponding to kernels of increasing complexity are equal to each other, the regularization results only in a sparse bottleneck and in a network preference to use a single kernel. There is no preference towards simpler kernels yet. We see that for 1D TFIM in $z$ basis, TetrisCNN tends to use kernel (3,1) with dilatation 1 accompanied by kernel (2,1) with dilatation 1. For 1D TFIM in $y$ basis, the variety of used kernels is even larger, the dominant one also being kernel (3,1) with dilatation 1, but accompanied by kernels (3,1) with dilatation 2 and 3.

\paragraph{Non-uniform $\lambda_k$} In the case of 1D TFIM in $z$ basis, for a bit larger differences between $\lambda_k$ ($\lambda_{\rm max} \geq 10^{-3}$), the dominant kernel changes to (2,1), and then to kernel (1,1) for $\lambda_{\rm max} > 10^{-1}$. These choices stay the same across 5 initializations. This kernel switch results only in a slight $R^2$ drop, suggesting that (1,1) learns a very relevant spin correlator. For $y$ basis, starting from $\lambda_{\rm max} = 10^{-2}$, all network instances use kernel (2,1). From $\lambda_{\rm max} \geq 10^3$, the kernel choice becomes unstable and is accompanied by a large $R^2$ drop, suggesting that using smaller kernels is detrimental. 

\paragraph{TetrisCNN trained on 2D IGT is less dependant on $\lambda_k$} Finally, TetrisCNNs trained on 2D IGT samples, regardless of $\lambda_k$, usually use kernel (2,2) as the dominant kernel. For $\lambda_{\rm max} \leq 10^{-2}$, the only difference is that the (2,2) kernel is accompanied by some larger kernels, and the choice becomes less stable across initializations. Interestingly, the largest $R^2$ is for intermediate $\lambda_{\rm max} = 10^0$ suggesting that sparsity does not always come at the cost of performance drop.

\paragraph{The $R^2$ difference between TetrisCNNs trained on 1D TFIM snapshots from $z$ and $y$ basis} Next to the discussion on $\lambda_k$ and the use of kernels, the first row of \fig~\ref{fig:lambda_sparsity}(a)-(b) nicely suggests via the $R^2$ difference that the $z$ basis contains better information for predicting the tuning parameter $g$ of 1D TFIM than the $y$ basis. Indeed, the $g$ aims to align spins in the $z$ direction, and it is the term breaking the symmetry underlying the phase transition. Following the regression performance between measurement bases is, next to the simplicity of the learned spin correlator, an important guidance on studying phases of matter with TetrisCNN using various measurements.

\section{Symbolic regression results}\label{app:sr_results}
\begin{figure}[t]
    \centering
    \includegraphics[width=\linewidth]{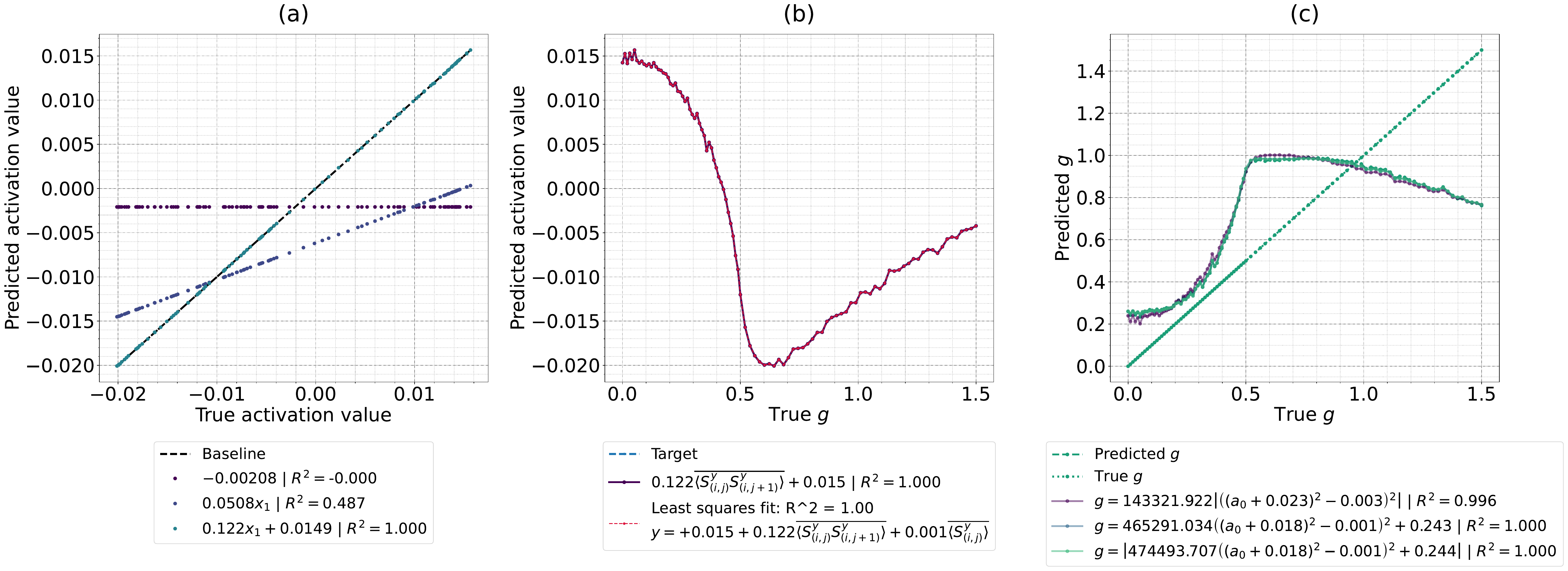}
    \caption{\textbf{Symbolic regression results for TetrisCNN trained on 1D TFIM measured in $y$ basis.} (a) Fits for $a_{[(2,1),1]} (x_1)$, where $x_1 = \langle \overline{S^y_i S^y_{i+1}} \rangle$. (b) Fits for $g(a_{[(2,1),1]})$. (c) Fits for $g(a_0)$, where $a_0 = \langle \overline{S^y_i S^y_{i+1}} \rangle$ and comparison between the predicted $g$ of the network based on snapshots from the $y$ basis and the true $g$.}
    \label{fig:sr_TFIM_y}
\end{figure}

The analysis process described in App.~\ref{sec:app:interpretation_theory} is presented here on the example of 1D TFIM model in $y$ basis in Fig.~\ref{fig:sr_TFIM_y}. Panels (a) and (b) present the mapping between the spin correlators computed from the raw input data and the learned bottleneck activations as a function of the regression parameter. In all panels, the regression parameter is the transverse field strength $g$. In panel (b), we also show the least squares fit that includes all possible correlators from \eq~\eqref{eq:2_1_y}, i.e., $\langle \overline{S_i^y} \rangle$ and $\langle \overline{S_i^y S_{i+1}^y} \rangle$. The coefficient next to $\langle \overline{S_i^y S_{i+1}^y} \rangle$ is two orders of magnitude larger than the one next to $\langle \overline{S_i^y} \rangle$, indicating that the regularization of the bottleneck is working as planned.
Finally, panel (c) presents the mapping between the bottleneck activations and the output of the network.

In \seclab~\ref{sec:results} we present the distilled equation for TetrisCNN predicting transverse field strength $g$ in the TFIM model out of measurements in $y$ basis, but our analysis also spanned two other experimental settings: TFIM model in $z$ basis and IGT model. In both cases, the symbolic distillation yielded an accurate description of the knowledge extracted by the network. 

\textbf{TFIM model in $z$ basis.} Here, the network correctly picked up the simplest nonzero correlator present in the data, which is the magnetization in the direction $z$, using kernel (1,1). The nonzero branch corresponding to this correlator is approximated in \eq~\eqref{app:eq:tfi_z_corr}. Given this approximation, an equation that approximates the whole TetrisCNN is presented in \eq~\eqref{app:eq:tfi_z_output}. 
\begin{gather}\label{app:eq:tfi_z_corr}
a_{[(1,1),1]} = - 1.176 \langle \overline{ S_i^z} \rangle + 0.426 \quad {\scriptstyle(R^2 = 1.00)} \\ \label{app:eq:tfi_z_output}
g = | 2.798 \langle \overline{ S_i^z} \rangle - 1.013|  \quad {\scriptstyle(R^2 = 0.97)}
\end{gather}

\textbf{IGT model} Analysis of data from IGT model presented us with a new insight into case previously studied by \reflab~\cite{greplova_unsupervised_2020}. We knew from their investigation that the correlations picked up by the CNN must not exceed the receptive field of convolutional kernel (3,3), and using our analysis routine we were able to narrow this estimate from more than $250 000$ possibilities to just one meaningful correlation - a four body correlator within the convolutional kernel (2,2). Our analysis of the nonzero TetrisCNN branch revealed this kernel to be $\langle \overline{ S_{(i,j)}^1 S_{(i,j+1)}^1 S_{(i,j+1)}^2 S_{(i-1,j+1)}^2} \rangle$, which is a product of all spins about a given vertex, producing a pattern similar to the vertex operator present in definition of the toric code (see \fig~6 of \reflab~\cite{greplova_unsupervised_2020}). Consecutive investigation of the raw data revealed that this correlator is indeed the only one in the (2,2) receptive field that varies continuously across the investigated $\beta$ range. All other correlators of smaller or equal size converge to a constant value as the number of snapshots per $\beta$ is increased. An equation describing the nonzero branch activation is presented in \eq~\eqref{app:eq:igt_corr}. The full distilled relation describing learned mapping from the input correlators to the regression parameter, inverse temperature $\beta$, is presented in \eqs~\eqref{app:eq:igt_output_1} - \eqref{app:eq:igt_output_2}.

\begin{gather}\label{app:eq:igt_corr}
a_{[(2,2),1]} = 4.95 \langle \overline{ S_{(i,j)}^1 S_{(i,j+1)}^1 S_{(i,j+1)}^2 S_{(i-1,j+1)}^2} \rangle - 0.322 \,\, {\scriptstyle(R^2 = 0.99)} \\ \label{app:eq:igt_output_1}
\beta = 1074.214 (\langle \overline{ S_{(i,j)}^1 S_{(i,j+1)}^1 S_{(i,j+1)}^2 S_{(i-1,j+1)}^2} \rangle + \\ \label{app:eq:igt_output_2}
+ 0.002 | 6421 \langle \overline{S_{(i,j)}^1 S_{(i,j+1)}^1 S_{(i,j+1)}^2 S_{(i-1,j+1)}^2} \rangle - 392 | - 0.044 )^2 + 0.015  \,\, {\scriptstyle(R^2 = 0.99)}
\end{gather}



\end{document}

%% file: our_commands.tex
\newcommand{\fig}{Fig.}

\newcommand{\tab}{Tab.}
\newcommand{\app}{App.}

\newcommand{\seclab}{Sec.}

\newcommand{\eq}{Eq.}
\newcommand{\eqs}{Eqs.}
\newcommand{\reflab}{Ref.}
\newcommand{\reflabs}{Refs.}

\definecolor{green}{rgb}{.2,.6,.2}
\definecolor{brickred}{rgb}{0.8, 0.25, 0.33}
\definecolor{brightcerulean}{rgb}{0.11, 0.67, 0.84}
\definecolor{maroon}{rgb}{0.788, 0.0, 0.086}
\definecolor{ao}{rgb}{0.0, 0.5, 0.0}